\documentclass{appolb}
\usepackage{epsfig}
\usepackage{graphicx}
\begin{document}
\title{HYPERNUCLEI - THE NEXT DECADE%
\thanks{Supported by the Bundesministerium f\"ur Bildung und Forschung (bmb+f) under contract No. 06MZ9182 and by the European Community - Research Infrastructure Integrating Activity Study of Strongly Interacting Matter (HadronPhysics2, Grant Agreement No. 227431; SPHERE network) under the 7. Framework Programme of the European Union.}%
}
\author{Josef Pochodzalla
\address
Institut f\"ur Kernphysik\\ Johannes Gutenberg-Universit\"at Mainz\\
Johann-Joachim-Becher-Weg 45, D-55128 Mainz, Germany\\
E-mail: \em{pochodza@kph.uni-mainz.de}
}
\maketitle
\begin{abstract}
We are at the verge of a new impact from hypernuclear experiments planned or already operative at various laboratories all over the world. The complementary of these different experimental approaches to hypernuclei provides a wide basis for a comprehensive understanding of strange hadrons in cold hadronic matter. High precision studies of light $\Lambda$ hypernuclei, spectroscopy of double $\Lambda\Lambda$ nuclei and the properties of antihyperons in nuclei are examples for the outstanding challenges for hypernuclei research in the next decade.

\end{abstract}

\section{Bridging the gap between quarks and stars}
In the early years of nuclear physics research the composition of an atomic nucleus in terms of protons and neutrons, its structure and basic properties were in the spotlight. Studies were focused on the nature of radioactive decays, nuclear reactions, and the synthesis of new elements and isotopes. Nowadays a nucleus is seen as a system of quarks and gluons that arrange themselves into protons and neutrons.  As a consequence the scope of nuclear science has broadened and extends from the today's fundamental particles - quarks and gluons - to the most spectacular of cosmic events like supernova explosions. Remnants of these cosmic catastrophes are neutron stars that have a core density about ten times higher than normal nuclei.
The properties of quarks and gluons are reasonable well understood and their mutual interaction is well described by the theory of Quantum Chromodynamics (QCD). But the appearance of neutrons and protons and other hadrons with their masses, charges, magnetisation and quark composition, together with the corresponding spatial distributions is not yet fully understood. How the nuclear force that binds protons and neutrons into stable nuclei or into neutron stars, emerges from QCD is yet another mystery and remains one of the greatest challenges for strong interaction physics.

In essence nuclear physics research attempts to understand the nature of all manifestations of nuclear matter in our universe - nuclei on the small scale and dense stellar objects on the large scale.
 Stable nuclei and neutron stars represent important checkpoints of the QCD phase diagram of cold baryonic matter. Such investigations are complementary to studies exploring the QCD phase diagram in highly dynamical, dense and hot quark-gluon matter created in ultra-relativistic heavy ion collisions, thus mimicking the early stage of our universe. Strangeness physics is adding a new degree of freedom to our understanding of hadrons, their structure, their interactions and the cooperative effects in the many-body environment in nuclear systems. In perspective, strangeness physics might be a cornerstone for further extensions into the regions of charm and potentially even higher flavors. Hence, strangeness physics might be well considered as being the gateway into flavor physics.

On the astrophysical scale the appearance of hyperons in the dense core of a neutron star has been a subject of extensive studies
since the early days of neutron star research \cite{Cam59}. It seems that irrespective of the hyperon-nucleon interactions,incompressibility, and symmetry parameter used, hyperons will appear in neutron stars at densities around 2-3 times normal nuclear density
\cite{Dap10}
and that the type of hyperons which dominates depends on the hyperon-nucleon interactions \cite{Gle01}. The additional strangeness degree of freedom softens the equation-of-state (EOS) leading to a smaller maximum masses of a neutron star compared to a purely nucleonic EOS. As a consequence the recent observation of a neutron star with about twice the solar mass \cite{NS10} rules out a large number of EOS's, particularly those involving hyperons. Nonetheless, there are still several high-density equations-of-states conceivable which allow neutron stars with masses close to or even beyond two times the solar mass \cite{Tak04,Rea09}. In these cases the origin of the extra repulsion which is needed to stiffen the EOS at high densities could  be related for example to a extra repulsion \cite{Tak04} similar to the three-body repulsion in conventional nuclear systems \cite{Wir93} or to nonlinear vector meson couplings \cite{Bed09}.

For many of these open questions hypernuclei can give authoritative answers or serve, at least, as laboratories for explorative studies. Hypernuclei are unique in their potential of improving our knowledge on the strange particle-nucleus interaction in a many-body environment and under the controlled conditions of a cold and equilibrated host system. This, in turn, is essential to derive eventually a more general and self-consistent description of the baryon-baryon interaction. Promising candidates are chiral effective field theory ($\chi$EFT), providing a link to QCD by obeying the relevant conservation laws and symmetries in the low energy-momentum domain of nuclear few-body systems. A unique feature of $\chi$EFT is their ability to determine in an order-by-order manner a hierarchy of many-body interactions which otherwise are hardly accessible in a systematic way. Thus nucleon-nucleon interactions from $\chi$EFT serve as a solid ground to link the basic long range nucleon-nucleon interaction to spectroscopic information of bound nuclei or NN scattering data. Because of the relatively large value of the strange-quark mass the extension of $\chi$EFT into the SU(3) sector is challenging and not yet fully solved. However, non-perturbative coupled channel approaches based on chiral SU(3) dynamics may offer a practical solution. While $\chi$EFT is at present primarily applicable to few body systems, other approaches -- reaching from relativistic and non-relativistic meson exchange models for free space YN and YY interactions up to nuclear density functional theory and the hypernuclear shell model -- are well suited for the description of medium and heavy mass hypernuclei and eventually neutron stars. Clearly these calculations need guidance by experimental data. High precision data on light hypernuclei are therefore of vital importance for the accurate determination of the missing parts of the YN and YY interactions.

\section{Future challenges for strangeness nuclear physics}
% fig.1 --------------------------------------------------------
\begin{figure}[t]
\begin{center}
\includegraphics[width=1.00\linewidth]{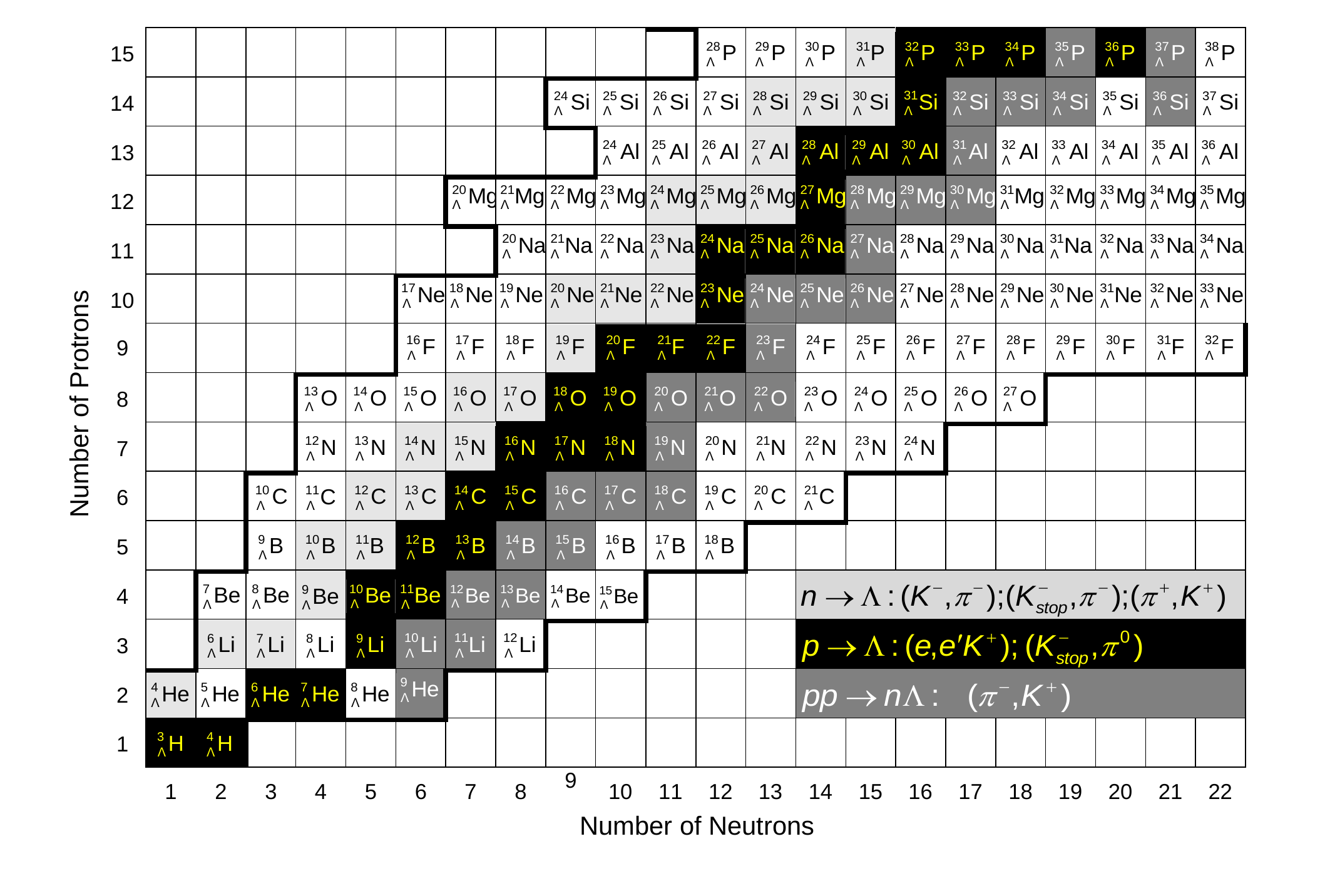}
\end{center}
\caption{Hypernuclei accessible by missing mass experiments for the different production channels. The boundaries at the neutron and proton rich side mark the predicted drip lines by a nuclear mass formula extended to strange nuclei.
}
\label{fig:zakopane01}
\end{figure}
% --------------------------------------------------------------

One can distinguish between two classes of reactions to produce and identify hypernuclei. One which relies on the detection of the decay products, the other which employs kinematic information of the production process to identify the produced hypernucleus.

On one hand, hypernuclei can be formed as secondary particles emerging from more or less violent hadronic interactions. The formation of a hypernucleus is usually tagged by its delayed weak decay producing a secondary vertex. Spectroscopic information is obtained exclusively by analysing the {\em decay products}. Examples are cosmic ray
interactions in emulsions or in bubble chambers, the production of single hypernuclei in proton or heavy ion induced reactions or the formation of double hypernuclei after a conversion of a $\Xi$-hyperon into two $\Lambda$ particles. Observables are the binding energies of hypernuclear groundstates, their lifetime by decay in flight, the ratio between the non-mesonic and the mesonic decay probability and the level structure by $\gamma$-spectroscopy.

During the first two decades of hypernucleus research nuclear emulsions were the main source of information on hypernuclei.
Even today emulsion data represent for many nuclei the most precise information on the $\Lambda$-binding energy. Around the turn of the last century high-resolution $\gamma$-spectroscopy of hypernuclei with germanium detectors became the most important tool for decay studies. These measurements provide precise information on the level schemes of various nuclei and allowed to extract different spin-dependent components of the $\Lambda$-nucleon interaction \cite{Tam10}.

On the other hand, employing quasi two-body kinematics, ground and excited hypernuclear
states can be identified by a missing-mass analysis of the incident beam and the observed associated meson. Examples are
the (K$^-$,$\pi^-$) and the ($\pi^+$,K$^+$) reactions which convert a neutron into a $\Lambda$ hyperon, the ($\gamma$,K$^+$) and (e,e'K$^+$) reactions which convert a proton into a hyperon and the double charge exchange
reaction ($\pi^-$,K$^+$). Since these reactions require stable
target nuclei, the hypernuclei accessible by these reactions are
limited. Figure \ref{fig:zakopane01} nicely illustrates the complementarity of the various production mechanisms and the need to study hypernuclei with different reactions to e.g. explore the charge symmetry breaking in mirror nuclei.
Electromagnetic probes are able to produce hypernuclei
 (black background) that cannot be generated with hadronic probes
(light and dark grey backgrounds) and vice versa.

Today hypernuclei experiments have revealed a considerable amount of hypernuclear features such as that the $\Lambda$ particle essentially retains its identity in a nucleus, the extremely small spin-orbit strength, the appearance of non-mesonic weak decays hyperons or the importance of the coupling. The recent observation of hypernuclei and antihypernuclei in relativistic heavy ion collisions by the STAR collaboration \cite{STAR10} and the ongoing search for them by the ALICE experiment at LHC  open a new window towards the transition from a quark-gluon system to the common world of hadrons and complex nuclei.
During the next decade hypernucleus physics will face several challenges. Precision measurement of ground state masses, $\gamma$-spectroscopy of double hypernuclei and antihyperons in nuclei represent such highlights.

% fig.2 --------------------------------------------------------
\begin{figure}[t]
\begin{center}
\includegraphics[width=0.77\linewidth]{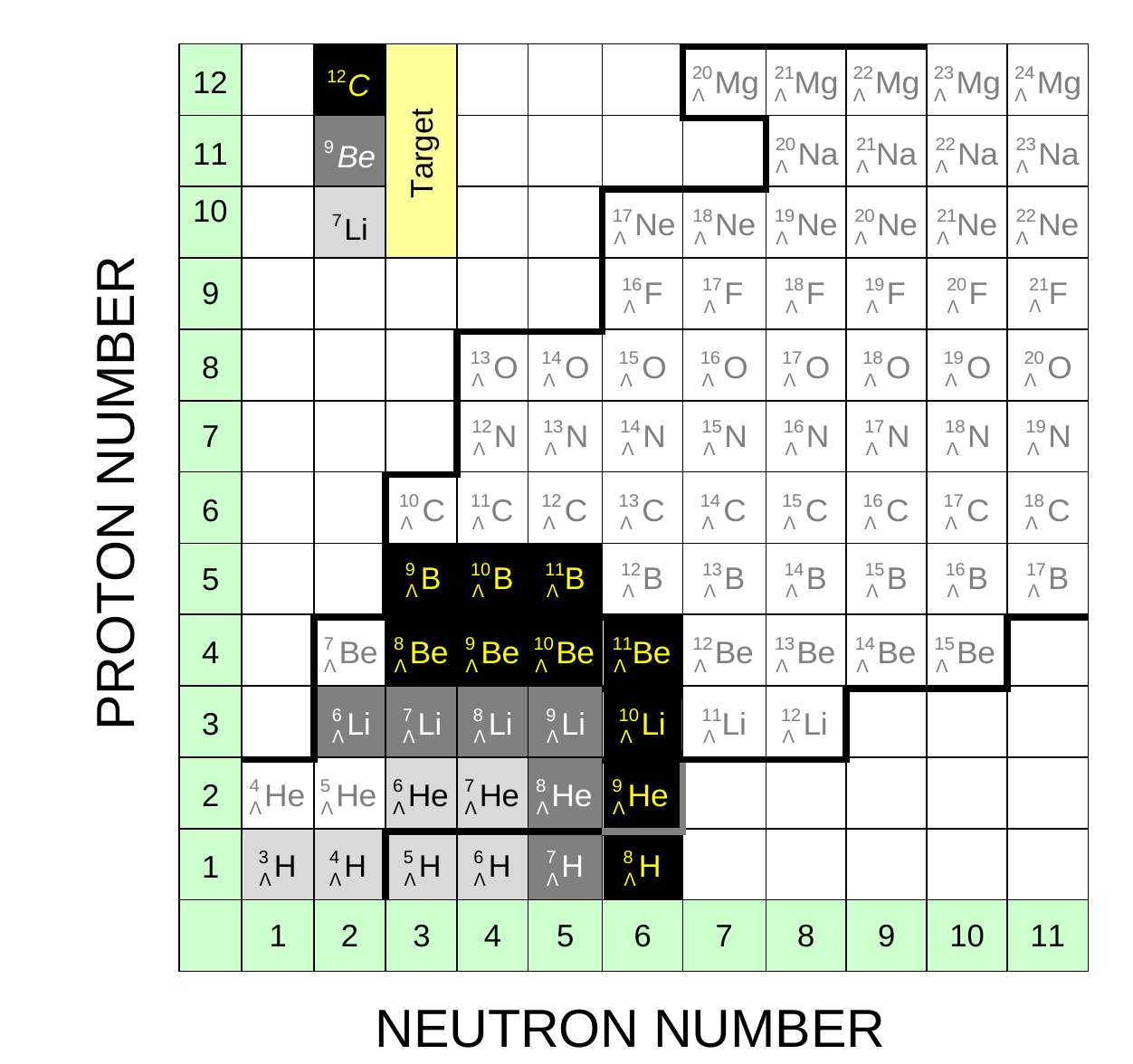}
\end{center}
\caption{Hypernuclei showing two-body pionic decays and which are accessible by high precision pion spectroscopy using a $^7$Li (light grey background), $^9$Be (grey) and $^{12}$C target (black). Already with $^7$Li and $^9$Be targets all isobars with A=6 can be studied.
}
\label{fig:zakopane02}
\end{figure}
% --------------------------------------------------------------
\subsection{Precision spectroscopy with electron beams}
Electro-production of hypernuclei offers the unique possibility to extract detailed structure information of ground and excited hypernuclear states. Confronting unique structure information on light $\Lambda$ nuclei with {\em ab initio} calculation for light nuclei will help to accurately determine the YN interactions and e.g. the role of three-body forces.  On the long term this will pave the road for planned investigations on $\Lambda\Lambda$ hypernuclei at FAIR to pin down the hyperon-hyperon interaction.

The monoenergetic pions produced in the weak mesonic two-body decay of hypernuclei are a direct measure of the $\Lambda$ binding energy. While missing mass experiments  are limited to nuclei close to the initial target (see Fig. \ref{fig:zakopane01}), the quasi-free kaon production of an excited primary hypernucleus - e.g. $^{12}C(e,e´K^+)^{12}B^*$ - and its subsequent decay gives access to a  variety of light and exotic hypernuclei, some of which cannot be produced or measured precisely by other means (Fig. \ref{fig:zakopane02}).
An instrument of central importance for the strangeness program at MAMI is the newly installed magnetic double spectrometer KAOS. Its compact design and its capability to detect negative and positive charged particles up to the highest particle momenta simultaneously significantly extends the capability of the spectrometer facility.
Pions from the weak decay from the secondary hypernuclei will be detected by one of the high resolution spectrometers of the A1 collaboration. It's momentum resolution ($<10^{-4}$) ultimately limits the reachable binding energy resolution to better than 10 keV. The high energy definition of the MAMI beam will be essential for a precise absolute momentum calibration of the high resolution $\pi$-spectrometer.

For example, already with a $^ 7$Li target the binding energies of $^3_{\Lambda}$H, $^4_{\Lambda}$H, $^5_{\Lambda}$H, $^6_{\Lambda}$H, $^6_{\Lambda}$He and $^7_{\Lambda}$He can be measured simultaneously (Fig. \ref{fig:zakopane02}). The $^4_{\Lambda}$H is particularly important because together with the $^4_{\Lambda}$He nucleus it represents a benchmark for our understanding of the charge symmetry breaking of the  $\Lambda$-N interaction. Employing step-by-step heavier targets will allow to map out a major part of the hypernuclear chart and may eventually provide precise binding energy information of extremely neutron rich hypernuclei. This new method will not only improve our knowledge on the binding energies of hypernuclei by about one order of magnitude, thus allowing for the first time precision studies of the charge symmetry breaking in hypernuclei.

Furthermore, electro-production of hypernuclei offers the unique possibility to vary the energy and momentum transfer independently. As a consequence the angular distribution of kaons associated with a given hypernuclear state is sensitive to the $\Lambda$-wave function inside the nucleus.

\subsection{Approaching the hyperon-hyperon interaction}
Double $\Lambda\Lambda$ hypernuclei represent a unique femto-laboratory to study the hyperon-hyperon interaction. Unfortunately the world supply of data on $\Lambda\Lambda$ hypernuclei is -- even half a century after the discovery of double hypernuclei -- very limited. Only few individual events have been identified so far \cite{HYPX}.

The simultaneous production and implementation of two $\Lambda$ particles into a nucleus is intricate.
The first observation of antihypernuclei by the STAR collaboration \cite{STAR10} impressively illustrates the potential to produce multi-strange hypernuclei in heavy ion
collisions \cite{Ker73,Bot07}. Unfortunately, high
resolution spectroscopy of excited states will not be feasible in these reactions. To
produce double hypernuclei in a more `controlled' way the conversion of a captured $\Xi^-$ and a proton into two $\Lambda$ particles can be used.
Relatively low momentum $\Xi^-$ can be produced using antiproton beams in $\overline{p}
\rightarrow \Xi^- \overline{\Xi}^+$ or $\overline{p} \rightarrow
\Xi^- \overline{\Xi}^\circ$ reactions if this reactions happens in
a complex nucleus where the produced $\Xi^-$ can re-scatter
\cite{Poc04}. The advantage as compared to the kaon induced
$\Xi$ production is that antiprotons are stable and can be
retained in a storage ring thus allowing a rather high luminosity.

Because of the two-step mechanism, spectroscopic studies based on
two-body kinematics cannot
be performed for $\Lambda\Lambda$ hypernuclei and spectroscopic information can
only be obtained via their decay products. The kinetic
energies of weak decay products are sensitive to the binding
energies of the two $\Lambda$ hyperons. While the double pionic decay of light double hypernuclei can be used
as an effective experimental filter to reduce the background
the unique identification of hypernuclei groundstates only via their pionic decay is usually
hampered by the limited resolution. In future, PANDA at FAIR intends to produce double-hypernuclei by numbers and study their high resolution $\gamma$-spectroscopy thus providing precise information on the level structure of these nuclei.

\subsection{Tackling cold, superdense hadronic matter}

The fundamental problem of the EOS of extremely dense strongly interacting matter constitutes the main enigma of neutron stars. On the long term Lattice QCD promises to revolutionize not only our understanding of isolated hadrons but also of our picture of multi-hadron systems and nuclei. First calculations of nucleon-nucleon and nucleon-hyperon interactions in the framework of quenched lattice QCD calculations were already performed. Eventually they may provide stringent boundary conditions for baryon-baryon potentials and will offer the extension into the regime of small distances respectively high densities.

Experimentally, the behavior of antihyperons in nuclei may provide new insights into the baryon-baryon interaction at extreme densities resp. short distances \cite{Poc08}.
Taking G-parity transformation \cite{Lee56} as a guidance a direct comparison of
the interactions of baryons with that of antibaryons in nuclei may help to shed
light on the nature of short-range baryon-baryon forces \cite{Duer56a}.
It is of course obvious that G-parity can only establish a link between the $NN$ and $N\overline{N}$ interactions for distances where meson exchange is a valid concept~\cite{Dov80,Fae82}. For distances smaller than about 1~fm, quark degrees of freedom may play a decisive role.

Concerning antibaryons, reliable information on their nuclear potential are available only for antiprotons. Antihyperons annihilate quickly in normal nuclei and spectroscopic information is therefore not directly accessible. In future quantitative information on the antihyperon potentials %relative to that of the corresponding hyperon
may be obtained via exclusive antihyperon-hyperon pair production close to threshold in
antiproton-nucleus interactions \cite{Poc08}. Once these hyperons
leave the nucleus and are detected, their asymptotic momentum
distributions will reflect the depth of the respective potentials.
In Ref.~\cite{Poc08} it was demonstrated that momentum correlations
of coincident hyperon-antihyperon pairs can be used to extract
information on the relative potential of hyperons and antihyperons in nuclei.

\section{Worldwide network of hypernuclear experiments}
% fig.4 --------------------------------------------------------
\begin{figure}[t]
\begin{center}
\includegraphics[width=1.0\linewidth]{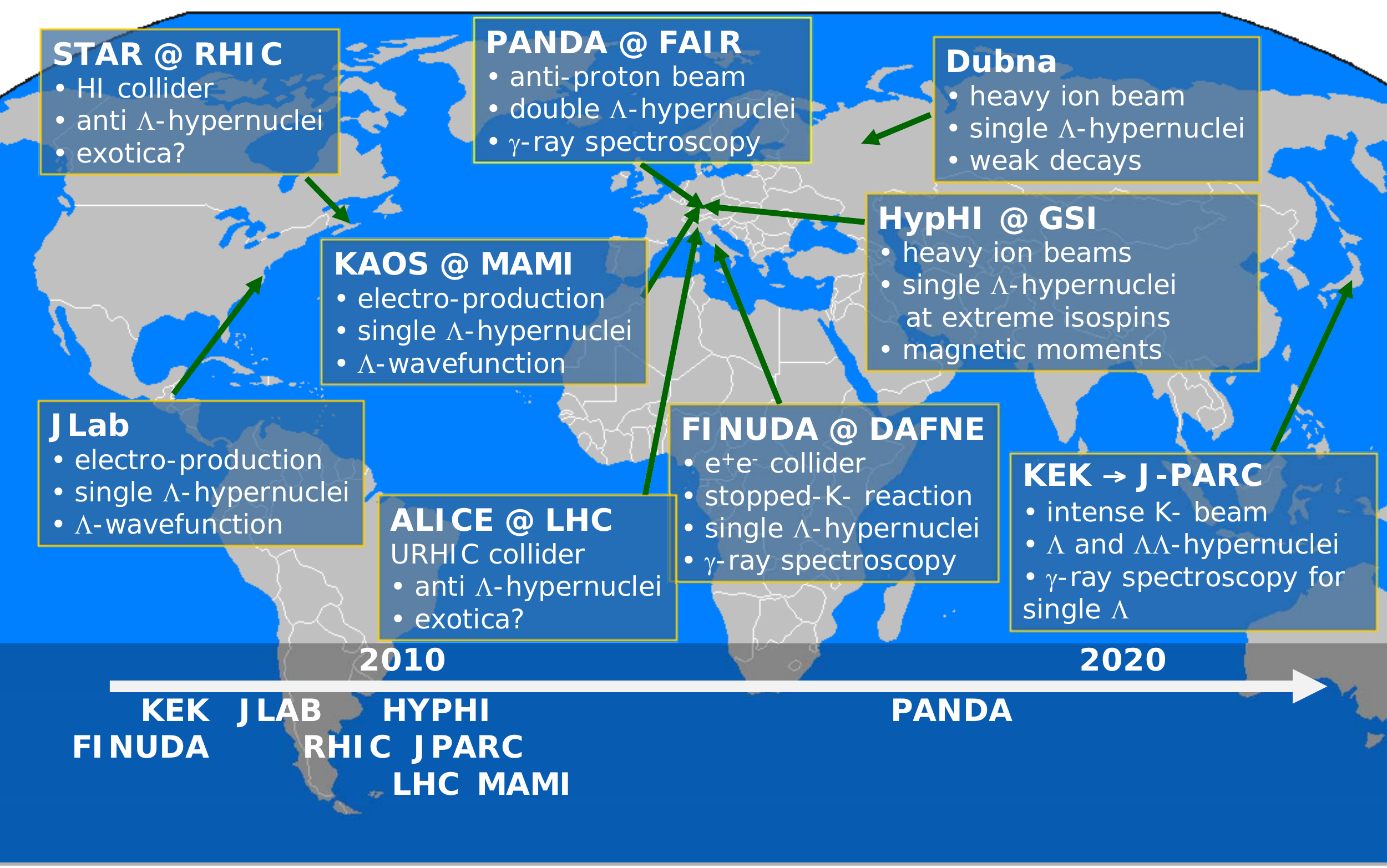}
\end{center}
\caption{Present and planned hypernuclear activities.
}
\label{fig:zakopane03}
\end{figure}
% --------------------------------------------------------------

Today we are at the verge of a new impact from the experiments planned or already operative at various laboratories all over the world (Fig. \ref{fig:zakopane03}). The complementary of these different experimental approaches to hypernuclei provides a wide basis for a comprehensive understanding of strange hadrons in cold hadronic matter.
\begin{itemize}
\item
At J-PARC hypernuclei will be produced by intense meson beams. Indeed hypernuclear physics is the main topic in several day-one experiments. The experiments will focus on the $\gamma$-spectroscopy of single $\Lambda$-hypernuclei, $\Xi$-hypernuclei by the missing mass method and ground state decays of double hypernuclei in hybrid-emulsion experiments.

\item
The HYPHI experiment at GSI is searching for hypernuclei in peripheral heavy ion collisions at energies around 2AGeV. Final results of the first experiment are expected early 2011.

\item
The STAR collaboration has recently reported the first observation of hypernuclei and anti-hypernuclei in relativistic heavy ion collisions. Also the FOPI collaboration has presented first preliminary results on the search hypernuclei. These results and the ongoing search for hypernuclei by the ALICE experiment at LHC open a new window towards the transition from a quark-gluon system to the common world of hadrons and complex nuclei. Heavy ion experiments will provide precise information on the lifetime on single hypernuclei. The possibility to produce multistrange nuclei and to measure their groundstate masses needs however further studies.

\item
The experimental program at JLAB has been completed and at the moment the 12GeV upgrade is going on. Several light hypernuclei have been studied by electro-production so far. The analysis of the latest experiments performed in 2009 is ongoing and new results are expected soon. So far the hypernuclear studies at JLAB are limited to the detection of ground and excited states. Except of very light $^3_{\Lambda}$H and $^4_{\Lambda}$H hypernuclei no angular distributions have been measured so far. The experimental program at JLAB is not expected to continue before 2012. The future program is will focus on precision pion spectroscopy and the excitation spectrum of medium heavy single hypernuclei.

\item
With the commissioning of the 1.6\~GeV electron beam of MAMI-C the study of hypernuclei has become possible in Mainz. Thus the activity at JLAB will be complemented by experiments planned at MAMI. The key elements of the experimental program at MAMI are the measurement of the angle distribution of the hypernuclei close to zero degree by missing mass studies and the precision pion spectroscopy.

\item
PANDA at FAIR will study double hypernuclei by high resolution $\gamma$-spectroscopy. This experiment complements measurements of ground state masses of double hypernuclei in emulsions at J-PARC or in heavy ion reactions. In addition, hyperon-antihyperon production in antiproton-nucleus collisions will help to explore the potential of antihyperons in nuclei.
\end{itemize}

To summarize, even though a number of new experimental techniques have been developed in the field of hypernuclear physics in the last decade, our knowledge is still limited to a small number of hypernuclei on or near the  $\beta$-stability line. However several activities which are ongoing or planned for the next decade will focus on different observables thus helping to overcome the present limitation.

\end{document}